# Taxation and the relationship between payments and time spent

by Christopher Mantzaris* & Prof. Dr. Ajda Fosner*

* = University of Primorska, Faculty of Management; CM's email: 68223065 [at] student.upr.si

**ABSTRACT (300 words):**
Tax work is costly for society: Administrative tax labour is typically to a high degree shuffled off the government and onto every taxpayer by law. The higher the burden of any tax system, the costlier for society, as taxpayers are unable to engage in proper wealth creation when being kept busy with administrative tax work. This research finds evidence for a relationship between hours spent to comply with taxes and amount of tax payment. These findings help better understand tax administrative costs and ultimately may help reduce them. PwC and World Bank's final "Paying taxes"-publication (2019) contains tax data for most of the world's jurisdictions, in particular annual hours spent to comply with tax obligations (X) and annual amount of tax payments (Y), both for the year 2019. X and Y were plotted in 6 tests. A positive slope, satisfying p and r values, high mutual information and finally a conclusive scatter plot picture were the 5 requirements that all needed to be met to confirm a positive relationship between X and Y. The first 2 tests did not make any adjustments to the data, the next 2 tests removed cities – thereby avoiding the double counting of jurisdictions– and the final 2 tests removed cities *and* outliers. Each test pair uses for Y first *total number of payments*; and for each second test the *number of other payments*, which excludes income tax payments for profit and labour. All 5 requirements were met in every of the 6 tests, indicating a positive dependence. In addition, 4 confirmatory tests validate the methodology. The found relationship is noticeably stronger for the total number of tax payments. Findings indicate that taxpayers' time spent on tax, and thereby society's overall tax administrative costs, could be reduced by simplifying taxation processes, including tax collection and payments.

## INTRODUCTION

Understanding tax administrative burdens is necessary for reducing them. Most jurisdictions use self-assessment principles, where taxpayers must find out and declare their owed taxes themselves – or with the help of others, such as hired tax professionals (Okello 2014, p. 4). By demanding this, the government essentially outsources tax administrative burdens and –by law– transfers them onto its citizens, under penalty of punishment if not done correctly.

Therefore, Total tax administrative cost (Ttac) can be defined as: all costs –whether measured in currency, labour or other forms– for the revenue side of all taxes; for example for preparing documents, declaring or auditing (Mantzaris & Fošner 2024, 1).
Ttac therefore arises on the taxpayer side, then often called "compliance cost" and on the government side –which may be dubbed 'tax administrative cost in the narrow sense'– for example for auditing (Vaillancourt *et al.* 2013, vii). The taxpayer's side of costs –primarily arising from labour cost or time spent on tax– is due to the self-assessment principle usually significantly higher than the government's side (Vaillancourt *et al.* 2013, p. 97-98; Evans 2007, p. 457; Moody *et al.* 2005, pp. 2, 11). Hence investigating how exactly time spent on tax can be reduced is a promising starting point when aiming to reduce Ttac.
Society ideally reaches all the goals of taxation with as little resources –including time spent on it– as possible. Time unnecessarily spent on tax administration is better invested in value creation. This analysis contributes to public knowledge in the areas tax and tax administration and potentially helps policy makers improve taxation, by reducing time spent on it and thereby reducing Ttac.
A Literature Review follows this section, demonstrating the gap for this particular research area and explains how this paper fills it. The subsequent methodology section explains data, methods and interpretation. The Results section contains the graphs and interpretation of results to each graph. The paper ends with Discussion and Conclusions.

**Literature review, research necessity, gap and justification**

Publications converting hours spent into a monetary value –for quantifying the largest subcomponent of Ttac– are ample. For income tax, Blaufus *et al.* (2019, pp. 930-931) alone mention 6 different methods from 14 different sources who did just that, before producing a 15th such analysis. Vishnuhadevi (2021, Appendix) cites at least 7 other sources doing that for sales taxes (VAT/GST).
So a myriad of methods exist for arriving at a monetary value for hours spent on tax, which have been applied to many different tax jurisdictions. This paper does not aim to add to this area of research. It focuses instead on a more under-researched, closely related issue.

There is far less literature focusing on what causes hours spent on tax. One of those include Rasyid and Urumsah (2023, p. 52), who deem in their systematic literature review "socio economic conditions, tax characteristics, taxpayer characteristics, and the complexity of tax rules as drivers of high tax compliance costs" (quoted from the 2023-10-15 Google-translation). Smulders *et al.* (2016, p. 725) found that for businesses in South Africa, factors with "a significant influence" on "hours spent" on tax include turnover, which had a positive correlation, age of the business had a negative one; But also the legal form had an effect, with sole proprietors spending the most time. Eichfelder and Vaillancourt (2014, p. 29) identify the following drivers of tax compliance burdens as the most relevant:
1. Tax complexity,
2. Tax administration (includes authorities' customer orientation, e-government *etc.*),
3. Tax accounting requirements and
4. International tax questions.
It stands to argue that those findings remain rather general and hardly indicate concrete policy solutions.

The authors could not find any targeted analysis investigating the relationship between time taxpayers spend to comply with a tax system and the amount of tax payments they need to make a year. This paper contributes in filling that research gap.

It seems logical to assume higher amount of payments would tend to increase time spent to comply, because one tax payment is less complex, easier and hence should result in lower hours needed to comply. Though, this does not necessarily need to be the case, because for example a direct debit mandate for income tax prepayment results in the same amount of hours spent, regardless of how many payments are made. The amount of tax payments can either directly result in higher time spent, due to contributing themselves to higher overall tax complexity – so being a component of tax complexity. Or because they are an indicator for a tax system's complexity. Because it indicates how 'easy', 'purist', simple or low in complexity a jurisdiction wants to keep its overall tax system.
Though, none of the two options has to be the case. This paper will analyse whether there is evidence for at least one of the two theses being true.
Because of the theoretically possible 'debit mandate situation', the sub-column "Other taxes payments" – which excludes profit and labour tax payments– will be especially examined. Because it is more likely to contain information or be an indicator of the amount of other tax kinds, which may more likely result in or be an indicator of a more complicated or expensive-to-maintain tax system. This is, because most tax jurisdictions finance themselves mainly with income tax, of which profit and labour tax are the most important derivatives. Therefore the suspicion: If a system has more payments for other kinds of tax, it likely has more tax kinds other than income, which would make the system more complex and time-consuming to maintain.

# MATERIALS AND METHODS

## Materials
This paper analyses hours spent on tax and amount of annual tax payments for most of the world from data published in 2019Q4 by PwC and World Bank, referring to the year 2019. It is the largest data series of such kind and the latest available volume of that series. Lastly, it refers to the last pre-pandemic year, making it more comparable to non-pandemic times.

The inspected data is in particular
X: hours a taxpayer spends to comply and
Y: the amount of annual tax payments
for almost all of the world's tax jurisdictions, with N = 211, 189, 179 and 181 – depending on which data is excluded, such as outliers.

## Methods
The data points of each jurisdiction in the dataset are plotted, whereas the x-axis shows "Time to comply (hours)" and the y-axis is the amount of "Other tax payments" (Y1) or total "Number of payments" (Y2) respectively.
Adjustments of the raw data (Figure 1) for doubles or outliers are made as confirmatory tests (Figures 2-6). This is to examine whether doubles or outliners meaningfully distort results.
Finally, 4 confirmatory tests (Figures 7-10) test whether shuffling or randomly assigning values would destroy the relationship found, thereby testing the methodology.

Calculations and graphs result from the authors' Python-code, available at pastejustit.com/raw/ivjpvggaco and archived on 2024-12-07 at archive.md/BUF3A.

## Result requirements and interpretation
This paper defines thresholds for different metrics. Only where all those thresholds are passed, it will be regarded as evidence for the thesis (meaning evidence for a positive relationship between X and Y). Accordingly, when at least one of the thresholds is not met, it will be deemed evidence against the thesis.

R1: An upward slope, from a linear regression (indicating a positive relationship).
R2: $p<0.01$, with 'p' being the likelihood of the found results appearing if there was no positive correlation – the null-hypothesis is "there is no positive correlation" and p is the chance of the null-hypothesis being true (one-sided/one-tailed test, yet the results for a two-sided/two-tailed test are also shown, for a more illustrative, complete picture).
R3: $r>0.2$, with 'r' being the Pearson correlation coefficient and 'rr' being $r \cdot r$ or $r^2$.
r measures the degree of linear correlation, with 1 being perfect positive correlation, -1 being perfect negative correlation and 0 being absolutely no correlation;
R4: Mutual information at >0.5 of maximum mutual information.
R5: A conclusive scatter plot picture (subjective assessment necessary).

Maximum mutual information is defined as the mutual information at Y=X for each respective X-Y-set. The exact calculation for all values, including mutual information and maximum mutual information, can be found in the abovementioned Python code.
Mutual information may be a useful metric for issues that are neither linear or two-dimensional, nor Gaussian – it is deemed rather experimental. The latest 4 confirmatory tests (Figures 7-10) will show whether it is useful for this application. R5 is considered the most important metric, as it offers the most comprehensive information.

## RESULTS

The following Figures 1-6 present the results in a scatter plot graph. Above each figure, the adjustments made to the original data are described. The axes are labelled within the figures, whereas X is always "Time to comply (hours)" and Y is either Y1 ("Other taxes payments) or Y2 (total "Number of payments"). Below each figure, additional values –N, slope, p, r and mutual information– and the authors' interpretation in regards to R5, as well as a summary for the figure, are shown. Above *Result requirements and interpretation*-subsection fully explains R1 to R5.

### Figure 1: no adjustments, other payments

No adjustments to original data. Only Somalia removed, since X, Y1 and Y2 each contained the values "No Practice".

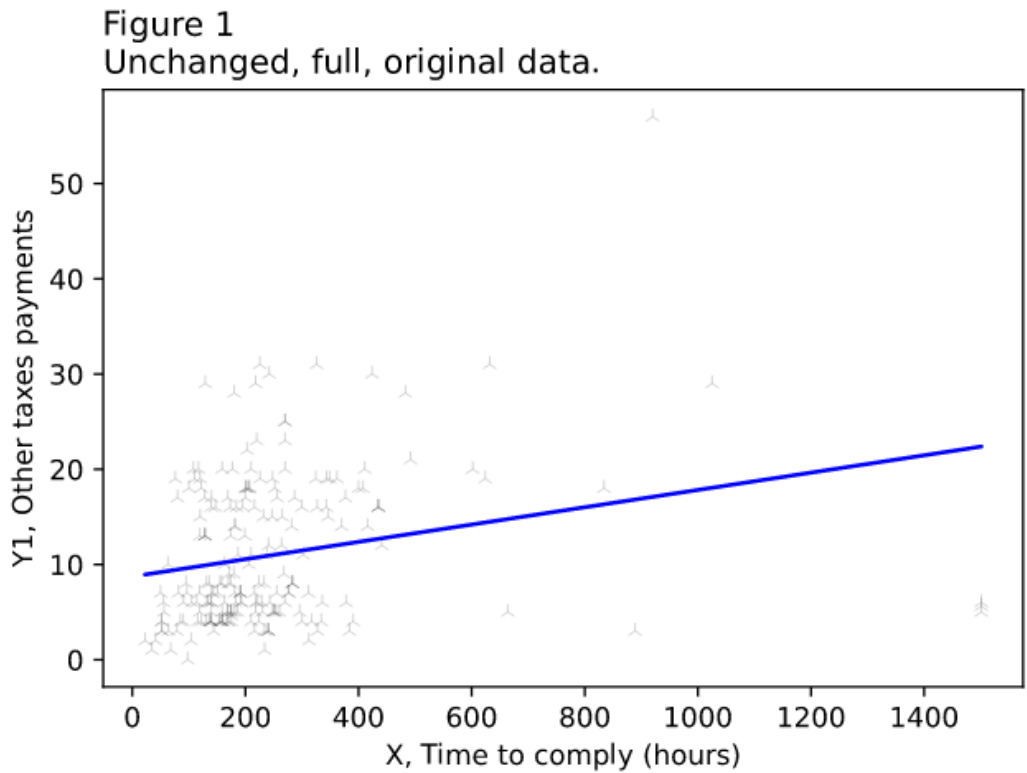

| | | | |
|---|---|---|---|
| N: 211 | | | |
| Slope: | 0.00908940652190793 | → >0 → | R1+ |
| p for no correlation (2sided/2tailed): | 0.00047521959389454376 | | |
| p for no positive correlation (1tailed): | 2p | → <0.01 → | R2+ |
| r: | 0.23849690353261718 | → >0.2 → | R3+ |
| rr: | 0.05688077299464651 | | |
| MutalInfo: | 2.758638948778665 | | |
| MaxMutInf: | 4.9214077605464235 | | |
| MutalInfo/MaxMutInf: | 0.5605385863154679 | → >0.5 → | R4+ |
| Scatter plot picture: | Indicates positive relationship | → | R5+ |

Summary for Figure 1: Results R1 to R5 all indicate a positive relationship.

Figure 1 shows a weak though statistically significant, positive dependency between time spent and amount of other tax payments. Doubles and outliers are noticeably present.

**Figure 2: no adjustments, total payments**
No adjustments to original data. Only Somalia removed, since X, Y1 and Y2 each contained the values "No Practice".

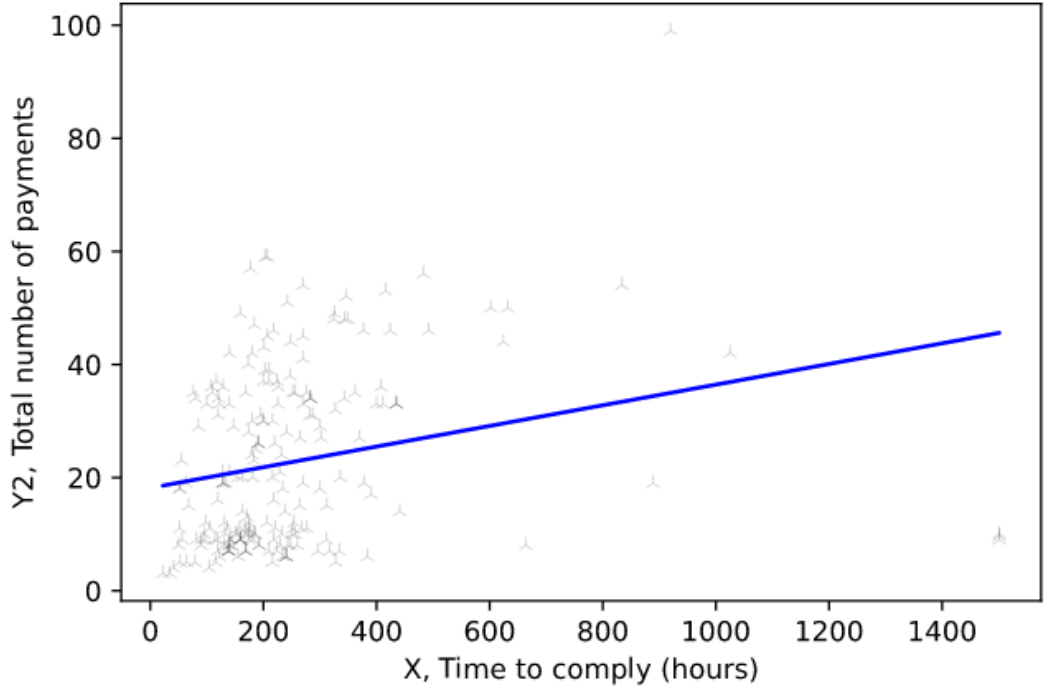

N: 211
| | | | |
|---|---|---|---|
| Slope: | 0.018259456351487954 | → >0 → | R1+ |
| p for no correlation (2sided/2tailed): | 0.0003638952057772288 | | |
| p for no positive correlation (1tailed): | 2p | → <0.01 → | R2+ |
| r: | 0.24317306397591293 | → >0.2 → | R3+ |
| rr: | 0.059133139043433446 | | |
| MutalInfo: | 3.3946340017407874 | | |
| MaxMutInf: | 4.9214077605464235 | | |
| MutalInfo/MaxMutInf: | 0.6897688968092905 | → >0.5 → | R4+ |
| Scatter plot picture: | Indicates positive relationship | → | R5+ |

Summary for Figure 2: Results R1 to R5 all indicate a positive relationship.

Figure 2 is similar to Figure 1, though total number of tax payments show a more pronounced, positive dependency with time to comply.

**Figure 3: removing cities, other payments**
Step 1: Somalia removed, since X, Y1 and Y2 each contained the values "No Practice".
Step 2: Removing cities –allowing only one x-y-pair per jurisdiction– namely: "Bangladesh - Chittagong", "Bangladesh - Dhaka", "Brazil - Rio de Janeiro", "Brazil - São Paulo", "China - Beijing", "China - Shanghai", "India - Delhi", "India - Mumbai", "Indonesia - Jakarta", "Indonesia - Surabaya", "Japan - Osaka", "Japan - Tokyo", "Mexico - Mexico City", "Mexico - Monterrey", "Nigeria - Kano", "Nigeria - Lagos", "Pakistan - Karachi", "Pakistan - Lahore", "Russian Federation - Moscow", "Russian Federation - Saint Petersburg", "United States - Los Angeles" and "United States - New York City".

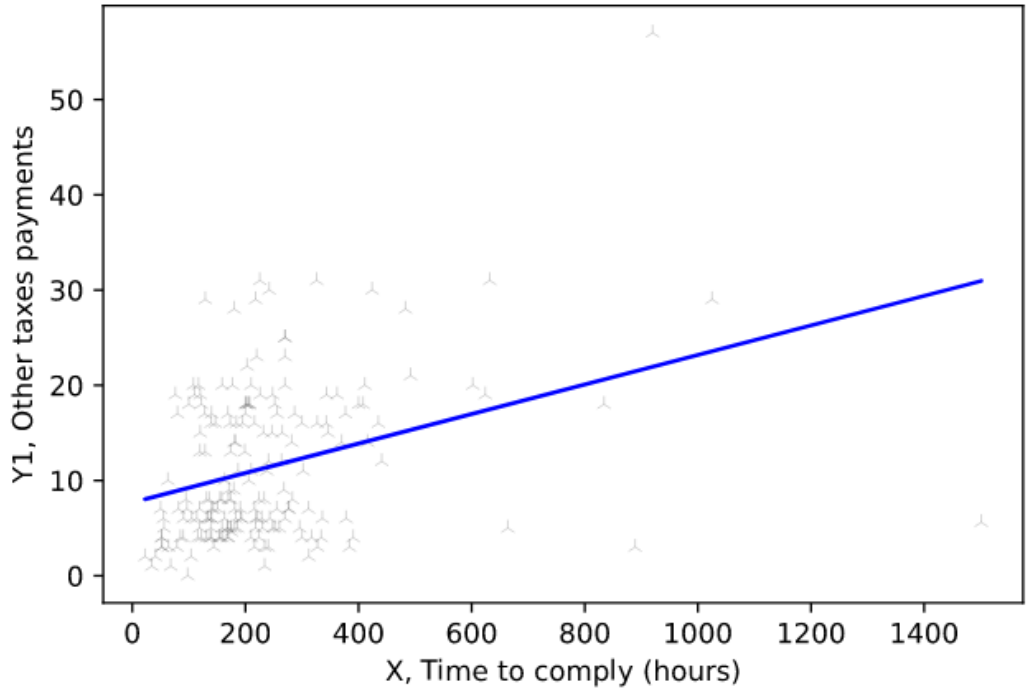

| | | | |
|---|---|---|---|
| N: 189 | | | |
| Slope: | 0.01550095323754606 | → >0 → | R1+ |
| p for no correlation (2sided/2tailed): | 2.375506061564407e-06 | | |
| Decimal form without exponents: | 0.000002375506061564407 | | |
| p for no positive correlation (1tailed): | 2p | → <0.01 → | R2+ |
| r: | 0.33545074495450666 | → >0.2 → | R3+ |
| rr: | 0.11252720229053348 | | |
| MutalInfo: | 2.810649601375341 | | |
| MaxMutInf: | 4.92583706234031 | | |
| MutalInfo/MaxMutInf: | 0.5705932952723319 | → >0.5 → | R4+ |
| Scatter plot picture: | Indicates positive relationship | → | R5+ |

Summary for Figure 3: Results R1 to R5 all indicate a positive relationship.

Removing doubles increases the lucidity of the positive relationship between X and Y1, compared to Figure 1.

**Figure 4: removing cities, total payments**
Step 1: Somalia removed, since X, Y1 and Y2 each contained the values "No Practice".
Step 2: Removing cities, allowing only one x-y-pair per jurisdiction.

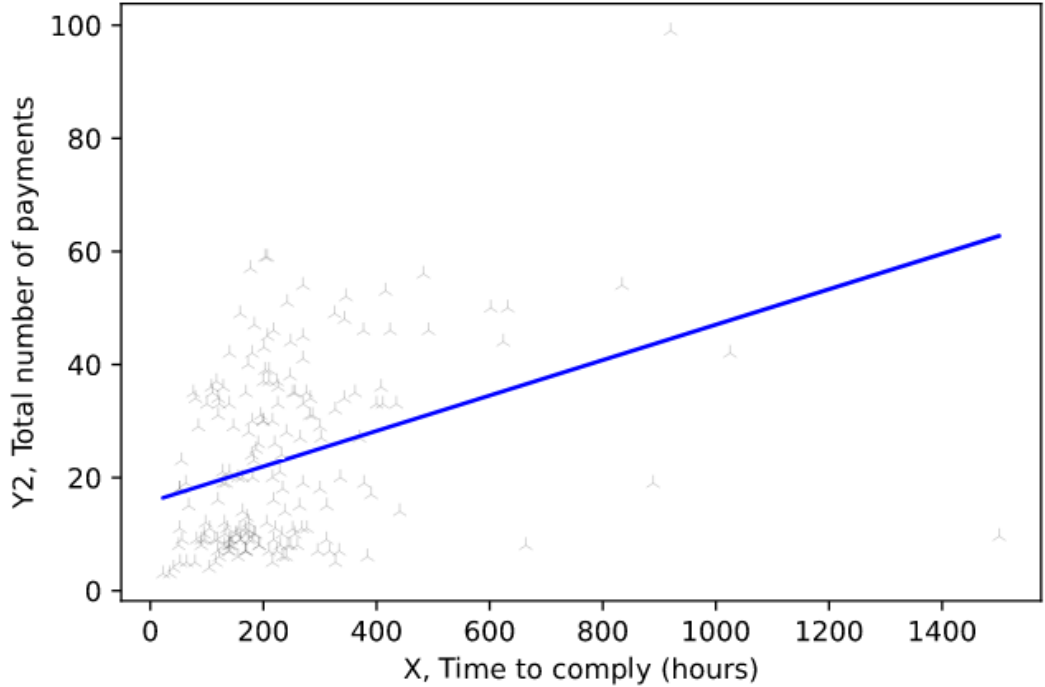

N: 189
| | | | |
|---|---|---|---|
| Slope: | 0.03130477878719239 | → >0 → | R1+ |
| p for no correlation (2sided/2tailed): | 9.505165312793962e-07 | | |
| Decimal form without exponents: | 0.0000009505165312793962 | | |
| p for no positive correlation (1tailed): | 2p | → <0.01 → | R2+ |
| r: | 0.3476613452692938 | → >0.2 → | R3+ |
| rr: | 0.12086841099445512 | | |
| MutalInfo: | 3.4353454485357315 | | |
| MaxMutInf: | 4.92583706234031 | | |
| MutalInfo/MaxMutInf: | 0.6974135370412694 | → >0.5 → | R4+ |
| Scatter plot picture: | Indicates positive relationship | → | R5+ |

Summary for Figure 4: Results R1 to R5 all indicate a positive relationship.

Similar to Figure 3, Figure 4 also depicts the dependency between X and Y2 clearer than Figure 2.

**Figure 5: removing cities then outliers, other payments**
Step 1: Somalia removed, since X, Y1 and Y2 each contained the values "No Practice".
Step 2: Removing cities, allowing only one x-y-pair per jurisdiction.
Step 3: Top and bottom 3 (≈1.6% of N) outliers of each X and Y1 removed, reducing N by a maximum of 12.

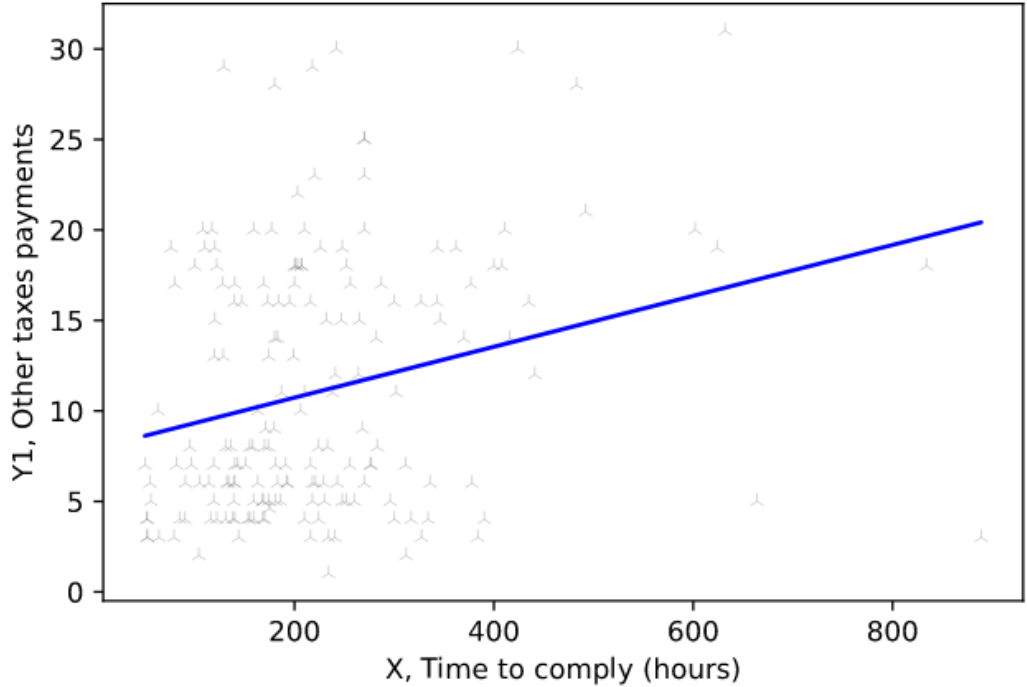

N: 179
| | | | |
|---|---|---|---|
| Slope: | 0.014071470001440682 | → >0 → | R1+ |
| p for no correlation (2sided/2tailed): | 0.00044852504837143475 | | |
| p for no positive correlation (1tailed): | 2p | → <0.01 → | R2+ |
| r: | 0.2596367832434706 | → >0.2 → | R3+ |
| rr: | 0.06741125921301694 | | |
| MutalInfo: | 2.6911997060302517 | | |
| MaxMutInf: | 4.853827252410844 | | |
| MutalInfo/MaxMutInf: | 0.5544490081911262 | → >0.5 → | R4+ |
| Scatter plot picture: | Indicates positive relationship | → | R5+ |

Summary for Figure 5: Results R1 to R5 all indicate a positive relationship.

Removing outliers decreases the statistical significancy across the board of R1 to R4 compared to Figure 3, though the overall scatter plot picture is little changed.

**Figure 6: removing cities then outliers, total payments**
Step 1: Somalia removed, since X, Y1 and Y2 each contained the values "No Practice".
Step 2: Removing cities, allowing only one x-y-pair per jurisdiction.
Step 3: Top and bottom 3 (≈1.6% of N) outliers of each X and Y2 removed, reducing N by a maximum of 12.

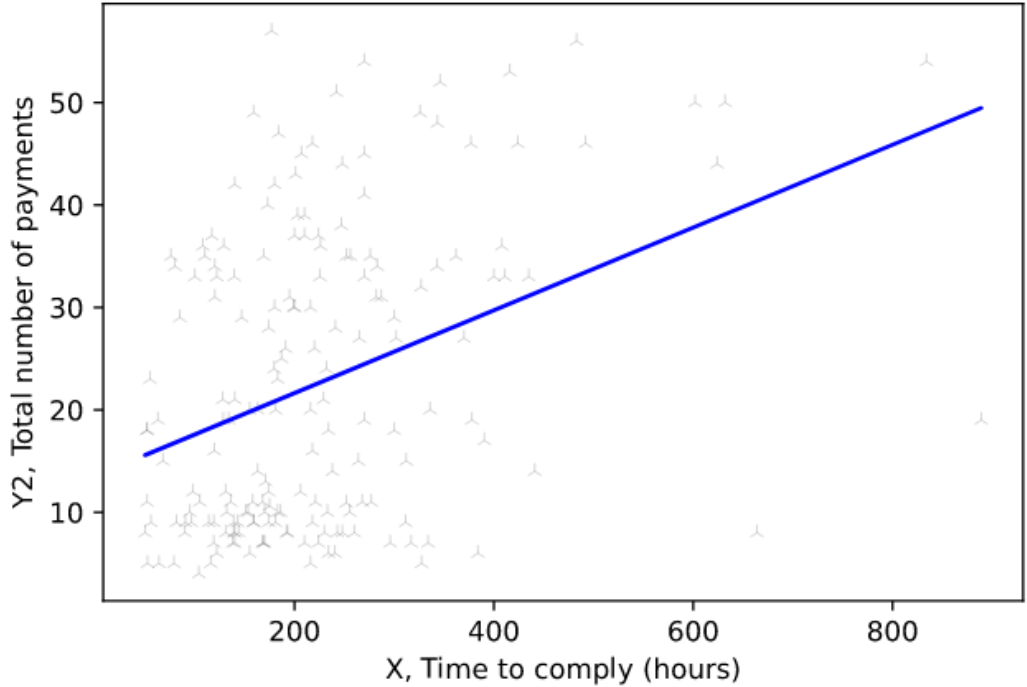

N: 181
| | | | |
|---|---|---|---|
| Slope: | 0.04038364915049969 | → >0 → | R1+ |
| p for no correlation (2sided/2tailed): | 4.5954524124023517e-07 | | |
| Decimal form without exponents: | 0.0000045954524124023517 | | |
| p for no positive correlation (1tailed): | 2p | → <0.01 → | R2+ |
| r: | 0.364363584665482 | → >0.2 → | R3+ |
| rr: | 0.13276082183027987 | | |
| MutalInfo: | 3.3610998944562396 | | |
| MaxMutInf: | 4.8839423774442015 | | |
| MutalInfo/MaxMutInf: | 0.688194011047101 | → >0.5 → | R4+ |
| Scatter plot picture: | Indicates positive relationship | → | R5+ |

Summary for Figure 6: Results R1 to R5 all indicate a positive relationship.

Figure 6 is across the board the most conclusive picture for statistical, positive dependency. Interestingly and in contrast to Figure 5, removing outliers resulted here in a stronger, positive relationship – shown across all metrics, including the scatter plot picture.

**Tests verifying the chosen methodology and result interpretation metrics**
To verify the results, four confirmatory tests were conducted. These tests assign X and Y values randomly. The posit is: Above values show a relationship, because X reveals information about Y (e.g. if X rises, Y is likely higher) and vice versa. Below confirmatory tests randomly assign X and Y and should therefore show no relationship. Because randomness means no relationship. If, however, random assignment would also indicate a relationship between X and Y, then above tests are faulty.
It is emphasised, that due to the random nature, reproducers will get slightly differing results. Reproducers may use the law of large numbers and conduct many tests, to verify that they did not receive an unlikely result, which could feign a different picture.

The four confirmatory tests are:
1. X and Y values are shuffled and randomly assigned to each other (Figure 7). This has the advantage, that the thereby created data will have the same probability distribution as the original data.
2. The corresponding Y value for every observed X-value was assigned with any random integer between the maximum and the minimum real, observed Y (Figure 8).
3. The corresponding X value for every observed Y-value was assigned with any random integer between the maximum and the minimum real, observed X (Figure 9).
4. X values are created with a random integer between the observed min and max X;
and Y values are created with a random integer between the observed min and max Y. These results are presented in Figure 10.

**Figure 7: based on Figure 6, shuffling X and Y**

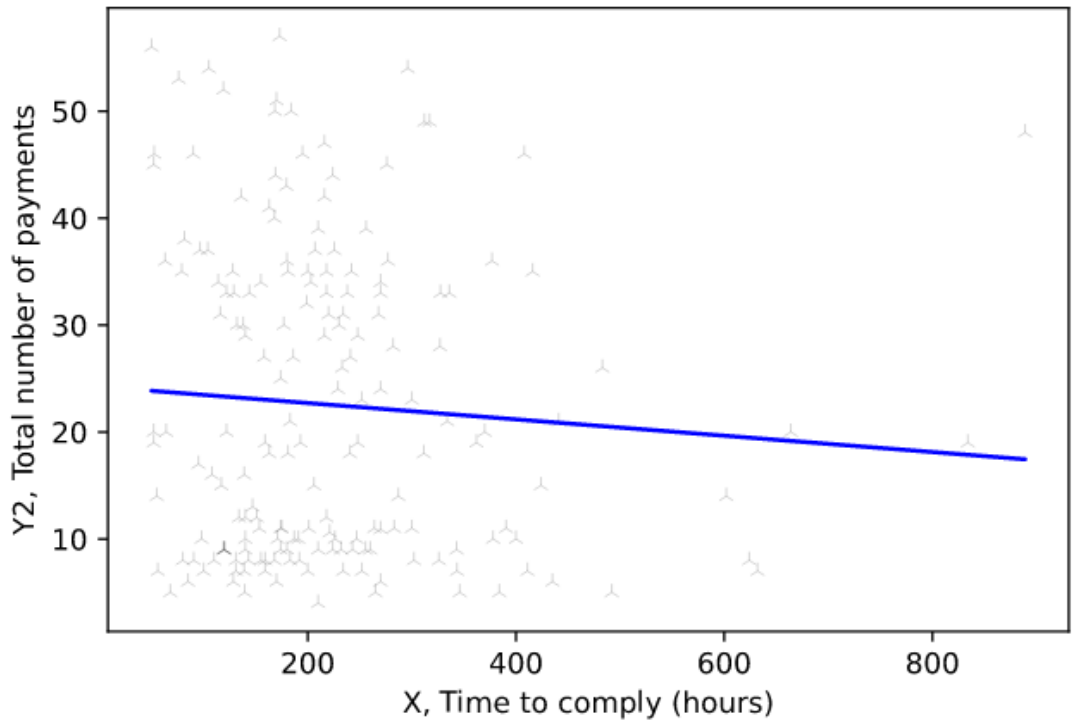

| | | | |
|---|---|---|---|
| N: | 181 | | |
| Slope: | -0.007634182310782464 | → ≤0 → | R1- |
| p for no correlation (2sided/2tailed): | 0.3568636973302423 | | |
| p for no positive correlation (1tailed): | 2p | → ≥0.01 → | R2- |
| r: | -0.06887980881519896 | → ≤0.2 → | R3- |
| rr: | 0.004744428062418361 | | |
| MutalInfo: | 3.353440809588174 | | |
| MaxMutInf: | 4.8839423774442015 | | |
| MutalInfo/MaxMutInf: | 0.6866257933499721 | → >0.5 → | R4+ |
| Scatter plot picture: | Indicates no relationship | → | R5- |

Summary for Figure 7: Results R1 to R5 do not all indicate a positive relationship. When values are shuffled, the positive relationship disappears, validating the applied methodology.
Interestingly, the mutual information values remain (nearly) identical.

**Figure 8: based on Figure 6, random X for every Y**

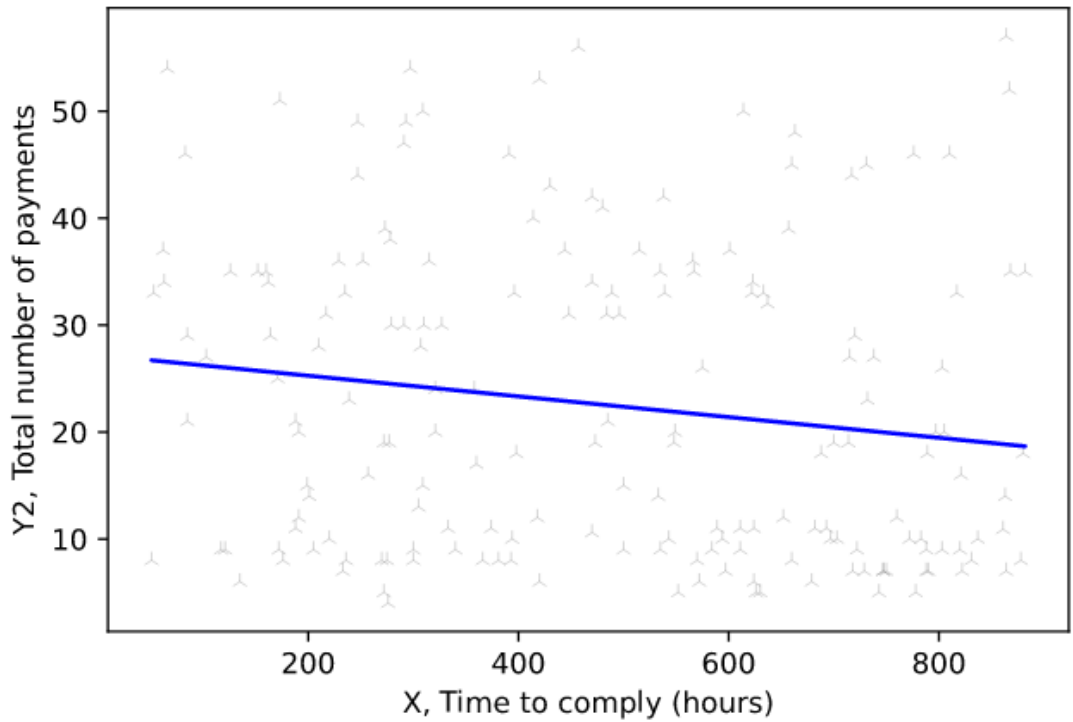

| | | | |
|---|---|---|---|
| N: | 181 | | |
| Slope: | -0.009675957773149061 | → ≤0 → | R1- |
| p for no correlation (2sided/2tailed): | 0.03346829332139968 | | |
| p for no positive correlation (1tailed): | 2p | → ≥0.01 → | R2- |
| r: | -0.1581551486190402 | → ≤0.2 → | R3- |
| rr: | 0.025013051034710694 | | |
| MutalInfo: | 3.504263807457964 | | |
| MaxMutInf: | 5.034765375313991 | | |
| MutalInfo/MaxMutInf: | 0.6960133285733145 | → >0.5 → | R4+ |
| Scatter plot picture: | Indicates no relationship | → | R5- |

Summary for Figure 8: Results R1 to R5 do not all indicate a positive relationship. When X values are randomly assigned, the positive relationship disappears, validating the applied methodology.

Notably, the mutual information values have slightly increased. This is somewhat surprising, given that only the distribution of the Y-values remained unaltered and X values are now distributed differently, namely randomly between min and max X. This could indicate that the distribution of Y values may also be close to a uniformly random distribution.

**Figure 9: based on Figure 6, random Y for every X**

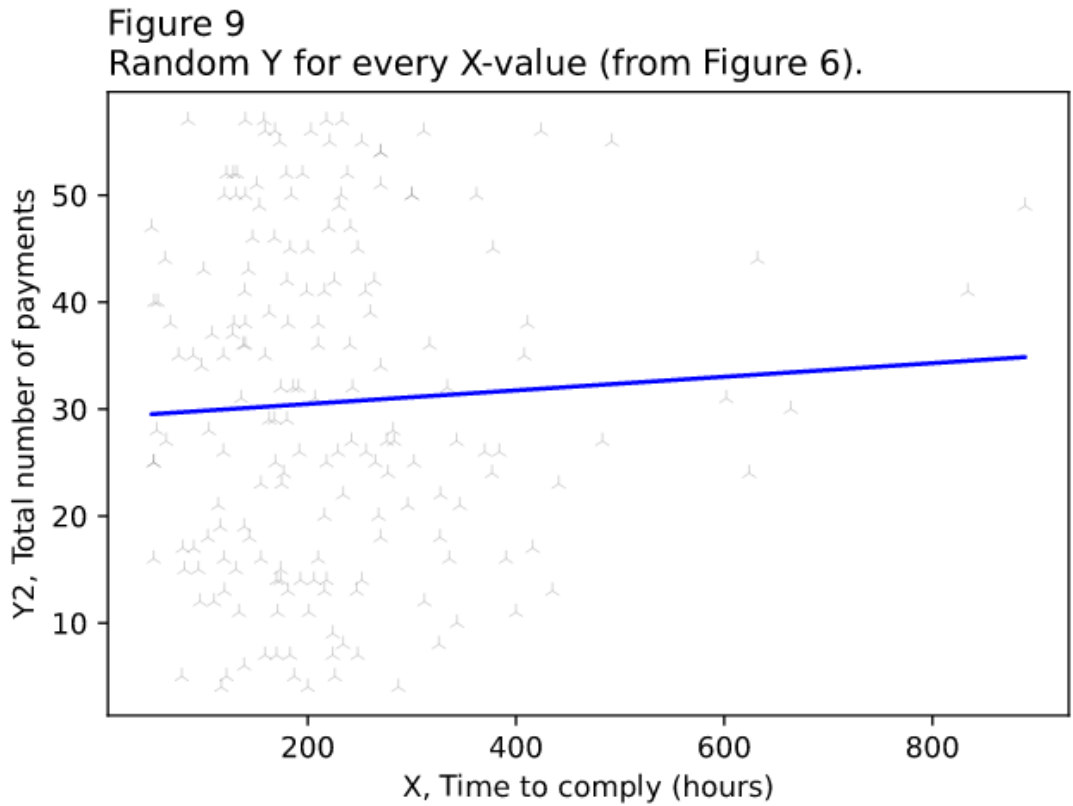

| | | | |
|---|---|---|---|
| N: | 181 | | |
| Slope: | 0.006357869958618561 | → >0 → | R1+ |
| p for no correlation (2sided/2tailed): | 0.47190978982942006 | | |
| p for no positive correlation (1tailed): | 2p | → ≥0.01 → | R2- |
| r: | 0.05380475287471598 | → ≤0.2 → | R3- |
| rr: | 0.0028949514319092573 | | |
| MutalInfo: | 3.5292443786210823 | | |
| MaxMutInf: | 4.8839423774442015 | | |
| MutalInfo/MaxMutInf: | 0.7226220347972161 | → >0.5 → | R4+ |
| Scatter plot picture: | Indicates no relationship | → | R5- |

Summary for Figure 9: Results R1 to R5 do not all indicate a positive relationship. When Y values are randomly assigned, the positive relationship disappears, validating the applied methodology. Remarkably, MutalInfo as percent of MaxMutInf has increased. This is somewhat surprising, given that the only the distribution of the X-values remains unaltered and Y values are now distributed differently, namely randomly between min Y and max Y. This perhaps indicates that the distribution of X values is also close to a uniformly random distribution.

**Figure 10: based on Figure 6, random X and random Y**

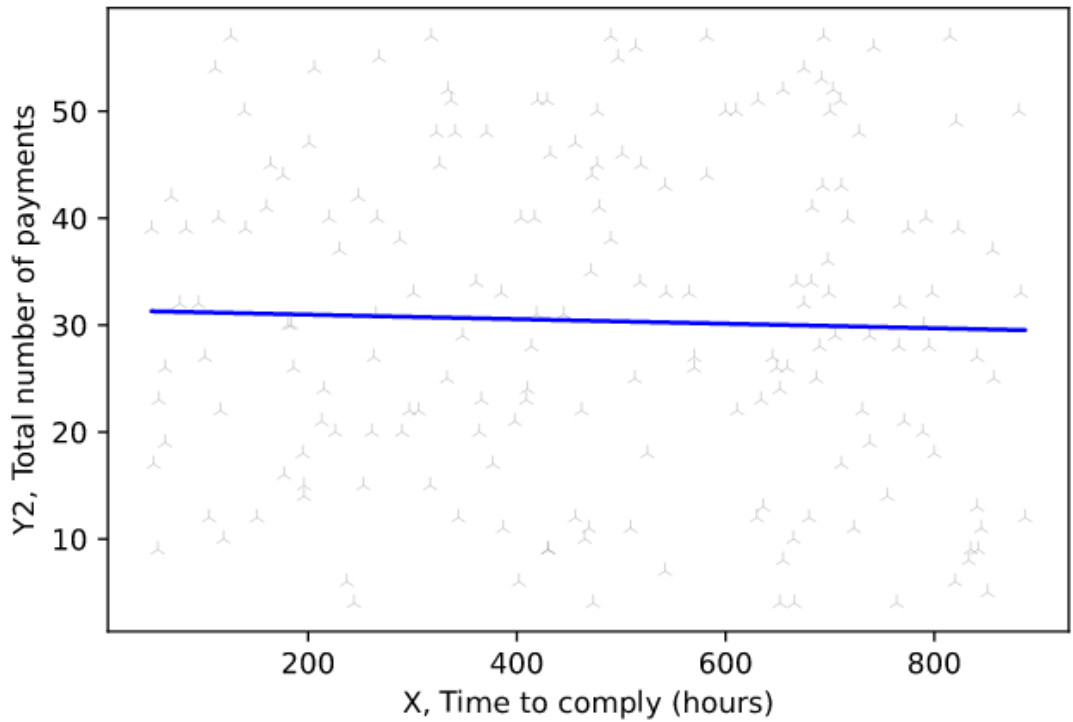

| | | | |
|---|---|---|---|
| N: | 181 | | |
| Slope: | -0.0021167523831243126 | → ≤0 → | R1- |
| p for no correlation (2sided/2tailed): | 0.6555764337534131 | | |
| p for no positive correlation (1tailed): | 2p | → ≥0.01 → | R2- |
| r: | -0.03337504803410965 | → ≤0.2 → | R3- |
| rr: | 0.0011138938312791264 | | |
| MutalInfo: | 3.759798389260027 | | |
| MaxMutInf: | 5.083610758244841 | | |
| MutalInfo/MaxMutInf: | 0.7395921064889178 | → >0.5 → | R4+ |
| Scatter plot picture: | Indicates no relationship | → | R5- |

Summary for Figure 10: Results R1 to R5 do not all indicate a positive relationship. When X and Y values are randomly generated, the positive relationship disappears, validating the applied methodology.

The mutual information values have increased, potentially because both X and Y show now the same probability distribution: A continuous, uniformly random assignment of values between their respective, originally observed min and max values.

**DISCUSSION**

Each of the Figures 1-6 shows in all 5 test results (R1 to R5) a relationship between hours spent on tax (X) and amount of tax payments (Y, as either: Y1, *other*; or Y2, *total number of tax payments*). The relationship between X and Y2 is notably stronger than that of X and Y1, in each R1 to R5, for each of the 3 pairs (= 6 tests in total). Overall, Figure 6 is most conclusive and illustrative in showing that Y2 and X

(namely *total number of tax payments* and *hours spent on tax*) share a statistically significant, positive dependence. The analysis adds indicative evidence to the thesis that higher tax complexity leads to increased administrative burden. In the current absence of similar research for this particular question, this paper contributes in filling the knowledge gap and informs policy makers, who aim for reducing society's tax costs.

Lastly, the tests validating the methodology revealed overall solidity in the methods and thresholds, yet mutual information as a percentage of maximal mutual information (R4) is not as useful of a metric in the pertinent analysis, compared to R1, R2, R3 or R5, in particular r and p values and the assessment of the scatter plot seem to be apt metrics.

**CONCLUSION**

Data for nearly all of the world's tax jurisdictions was examined in 6 tests; with N at 211, 189, 179 and 181, depending on which data was excluded, such as outliers. In each, all 5 thresholds were met. Those were R1 to R5, namely:
1) a positive slope,
2) $p<0.01$,
3) $r>0.2$, and
4) mutual information larger than 50% of maximum mutual information (MaxMutInf). MaxMutInf was defined as Y=X for each tested dataset. The 5th requirement was
5) a conclusive scatter plot picture.

These 5 thresholds are further explained in the *Result requirements and interpretation*-subsection. Methodology tests indicated mutual information to be less useful than hoped and the more traditional metrics (R1, R2, R3 and R5) to be more useful for this particular use case.

The results show: Indicative evidence was found to support
— the specific research question: that there is a relationship between X, hours spent to comply with tax duties (as a component of tax administrative cost), and Y, amount of tax payments (as an indicator or component of tax complexity);
— and hence also –albeit to a weaker degree– for the broader thesis: that tax complexity shows a dependency with higher costs for society.

This paper is the first investigation answering this specific research question. More research could help further fill this research gap and confirm, add to –or contest– the findings. Confirmatory tests underpinned results and methodology.

It should be noted, that the relationship found might be symptomatic rather than causal and that taxation is not a linear, two dimensional, but rather a multi-causal, complex issue. Nevertheless, those findings give direction for policy makers, as they indicate that reducing tax complexity –and in particular amount of payments– is likely to alleviate administrative burdens for society.

# NOTES ON APPENDICES

There is no appendix, other than the authors' Python code, found at pastejustit.com/raw/ivjpvggaco and archived on 2024-12-07 at archive.md/BUF3A.